\documentclass{aa}  

\usepackage{graphicx}
\usepackage{txfonts}
\begin{document}

   \title{Nonlinear internal waves breaking in stellar radiation zones}
   \subtitle{Parametrisation for the transport of angular momentum:\\ bridging geophysical to stellar fluid dynamics}

   \author{S. Mathis
          \inst{1}}

   \institute{Universit\'e Paris-Saclay, Universit\'e Paris Cit\'e, CEA, CNRS, AIM, F-91191 Gif-sur-Yvette, France\\
   \email{stephane.mathis@cea.fr}}

   \date{Received; accepted}

 
  \abstract
   {Internal gravity waves (hereafter IGWs) are one of the mechanisms that can play a key role to redistribute efficiently angular momentum in stars along their evolution. The study of IGWs is thus of major importance since space-based asteroseismology reveals a transport of angular momentum in stars, which is stronger by two orders of magnitude than the one predicted by stellar models ignoring their action or those of magnetic fields.}
   {IGWs trigger angular momentum transport when they are damped by heat or viscous diffusion, when they meet a critical layer where their phase velocity in the azimuthal direction equals the zonal wind or when they break. Theoretical prescriptions have been derived for the transport of angular momentum induced by IGWs because of their radiative and viscous dampings and of the critical layers they encounter along their propagation. However, none has been proposed for the transport of angular momentum triggered by their nonlinear breaking. In this work, we aim to derive such a physical and robust prescription, which can be implemented in stellar structure and evolution codes.}
   {We adapt an analytical saturation model, which has been developed for IGWs nonlinear convective breaking in the Earth atmosphere and has been successfully compared to in-situ measurements in the stratosphere, to the case of deep spherical stellar interiors. This allows us to derive the saturated amplitude of the velocity of IGWs breaking in stellar radiation zones through convective overturning of the stable stratification or the instability of the vertical shear of IGWs motion and of the angular momentum transport they trigger. In a first step, we neglect the modification of IGWs by the Coriolis acceleration and the Lorentz force, which are discussed and taken into account in a second step.}
   {We derive a complete semi-analytical prescription for the transport of angular momentum by IGWs, which takes into account both their radiative damping and their potential nonlinear breaking because of their convective and vertical shear instabilities. We show that the deposit of angular momentum by breaking waves is increasing with their latitudinal degree and the ratio of the Brunt-Va\"is\"al\"a frequency and the wave frequency and when the density decreases or the Doppler-shifted frequency vanishes. This allows us to bring the physical prescription for the interactions between IGWs and the differential rotation to the same level of realism that the one used in global circulation models for the atmosphere.}
   {}


   \maketitle
%
\section{Introduction}

Internal gravity waves (hereafter IGWs) are waves that propagate
in stably stratified stellar radiation zones. Their restoring force is the buoyancy force. Since more than 40 years, they have
been identified as one of the key mechanisms able to efficiently
transport angular momentum and chemicals in these regions \citep[e.g.][]{Press1981,Schatzman1993,Zahnetal1997}. Their
study is thus of major importance within the on-going probe of
the internal transport of angular momentum in the Sun and in
stars thanks to space-based helio- and asteroseismology \citep[e.g.][]{Garciaetal2007,Mosseretal2012,Mosseretal2024,Deheuvelsetal2014,Deheuvelsetal2015,Deheuvelsetal2020,VanReethetal2016,Aertsetal2017,Lietal2020, Lietal2024}. It reveals that the Solar and stellar interiors
are the seat of a strong extraction of angular momentum, more
intense by two orders of magnitude than what is predicted by
the stellar structure and evolution codes including the transport
mechanisms related to differential rotation \citep[e.g. vertical and
horizontal shear instabilities and meridional circulation;][]{Zahn1992,MaederZahn1998,MathisZahn2004,Mathisetal2018} but neglecting IGWs and magnetic
fields \citep{Marquesetal2013,Ceillieretal2013,Cantielloetal2014,Ouazzanietal2019}. In this framework, IGWs are with magnetic fields the
best candidate to explain the extraction of angular momentum
from the radiative core of solar-type stars and their internal mixing \citep[e.g.][]{MontalbanSchatzman2000,CharbonnelTalon2005,TalonCharbonnel2005}, in sub-giant stars \citep{Pinconetal2017}, and in early-type
stars \citep[][]{Rogers2015,Rogers2017,Neineretal2020,Vargheseetal2023}.\\

In this context, the key theoretical challenge is to provide the
more realistic and robust prescription as possible for the extraction or deposit of angular momentum by IGWs to be implemented
in stellar structure and evolution codes (hereafter SSE codes)
which compute the evolution of rotating stars \citep[e.g.][]{MaederMeynet2000,Palaciosetal2003,Marquesetal2013}. An IGW has three
channels to transport momentum and matter: through i) viscous
friction or heat diffusion \citep[e.g.][]{Press1981,Schatzman1993,Zahnetal1997,VadasFritts2005}, ii) critical layers, which
are the seat of a resonance between an IGW and a large-scale
(zonal) flow \citep[e.g.][]{BookerBretherton1967,Alvanetal2013}, and iii) nonlinear breaking because of IGWs convective
or shear instabilities \citep[e.g.][]{Lindzen1981,Sutherland2001,BarkerOgilvie2010,Barker2011}. If the physical description
of the transport of angular momentum by IGWs because of
their radiative damping has become the standard way of implementing
it in SSE \citep[e.g.][]{TalonCharbonnel2005,Fulleretal2014}, IGWs breaking in stars came on the forefront in several
configurations. First, \cite{Press1981} pointed out that progressive
IGWs can break near the center of solar-type stars because of
their geometrical focusing. This breaking is one of the main
sources of the dissipation of tidal gravity waves triggered by close-in
massive planets in solar-type stars \citep[e.g.][]{BarkerOgilvie2010,Barker2011}, which drives their orbital evolution \citep{Lazovik2021}. In early-type stars, IGWs excited by their convective core
can break near their surface because of the low-density there \citep[][]{Rogersetal2013,Rogers2015}.
In these configurations, specific physical modelling must be
developed to evaluate angular momentum transport triggered by
breaking IGWs while the standard formalism of linear IGWs
damped by radiative damping is not physically appropriate
anymore.\\

Such a theoretical modelling has been developed in the case
of the atmosphere of the Earth for convectivelly-excited IGWs \citep{Lindzen1981,HoltonZhu1984,Lottetal2012,LottGuez2013} to compute their impact on the general atmospheric circulation.
The derived prescriptions are then implemented in global circulation
models (GCM) where small-scale transport and mixing
processes are not solved directly but parameterised \citep[e.g.][]{Ribsteinetal2022,Achatzetal2024}, a situation
similar to SSE codes. The predictions of these prescriptions have been successfully compared to Reynolds stresses measured in the Earth atmosphere by stratospheric balloons \citep[][]{Lottetal2023}. However, this modelling for the Earth
has been build in local Cartesian shallow-water models, which are
not adapted to global spherical shells such that stellar radiative
zones. In this work, we revisit the formalism developed in
Geophysics to adapt it to the case of stars to provide the relevant
prescription for SSE codes. In \S 2, we recall the basics of the
mathematical modelling of the adiabatic and dissipative propagation of low-frequency
IGWs, which transport angular momentum. In \S 3, we detail the conditions for the convective and shear-induced breaking of IGWs and we derive the triggered vertical transport of energy and angular momentum. To obtain these results, we adapt the saturation scheme built in the case of the Earth atmosphere to the case of spherical thick stellar radiation zones. In \S 4, we discuss the impact of the Coriolis acceleration and of the Lorentz magnetic force on IGWs breaking. Finally, in \S 5, we present the conclusions and the perspectives
of this work.

\section{Propagation of low-frequency internal gravity waves}

IGWs that transport efficiently angular momentum are low-frequency progressive waves \citep[e.g.][]{Schatzman1993,Zahnetal1997}. To study their nonlinear breaking following the path of what has been done in atmospheric physics by \cite{Lindzen1981}, \cite{Lottetal2012} and \cite{LottGuez2013}, the first step is to recall the formalism to model their propagation.

\subsection{Adiabatic propagation}
We expand the macroscopic velocity in stellar radiation zones as:
\begin{equation}
{\boldsymbol V}\left(r,\theta,\varphi,t\right)=r\sin\theta{\overline\Omega}\left(r\right){\boldsymbol{\widehat e}}_{\varphi}+{\boldsymbol u}\left(r,\theta,\varphi,t\right),
\end{equation}
where $\overline\Omega$ is the rotation angular velocity, which is assumed here to be shellular \citep{Zahn1992}, and $\vec u$ is the velocity field of IGWs. We have introduced the spherical coordinates $\left\{r,\theta,\varphi\right\}$ and the corresponding unit vector basis $\left\{{\widehat{\boldsymbol e}}_j\right\}_{j=\left\{r,\theta,\varphi\right\}}$. 

We here adopt the so-called “shellular” rotation approximation, where the rotation is varying only (or mainly) with $r$ the radius, since our objective in this work is to provide prescriptions for state-of-the-art 1D stellar structure and evolution codes. These codes, when taking transport and mixing processes' parametrisations into account, are working in this theoretical framework since the work by \cite{Zahn1992} \citep[e.g.][]{Talonetal1997,MaederMeynet2000,Palaciosetal2003,TalonCharbonnel2005,Marquesetal2013,Charbonneletal2013,Mathisetal2013,Cantielloetal2014,Fulleretal2014}. This implicitly assumes that a strong transport of angular momentum is ongoing along the horizontal direction that enforces a weak latitudinal variation of the angular velocity in stellar radiation zones. In an hydrodynamical theoretical framework, this may be due to hydrodynamical shear instabilities of the horizontal differential rotation \citep[e.g.][]{Zahn1992,Maeder2003,Mathisetal2004,Mathisetal2018,Parketal2021} or the action of IGWs \citep[e.g.][]{Mathis2009}.  In an MHD theoretical framework, this may be due to MHD instabilities and potential resulting dynamo action \citep[e.g.][]{Spruit2002,Fulleretal2019,Petitdemangeetal2023}. The shellular approximation has also been strengthened by the results obtained by the helioseismic inversions of the rotation profile of the Sun that revealed a quasi-uniform rotation of the Solar radiative core until $0.25\,R_{\odot}$, where $R_{\odot}$ is the radius of the Sun \citep[e.g.][]{Thompsonetal2003,Garciaetal2007}. A similar trend is obtained by asteroseismic probe of the internal rotation of stars in the whole Hertzsprung-Russell diagram that reveals only a weak radial differential rotation in stellar radiative zones in agreement with what has been observed in the Solar case \citep[e.g.][and references therein]{Deheuvelsetal2012,Deheuvelsetal2014,Deheuvelsetal2015,Deheuvelsetal2020,Kurtzetal2014,Saioetal2015,Murphyetal2016,Aertsetal2017}.\\

We solve the linearised momentum equation
\begin{equation}
{\rm D}_{t}{\boldsymbol u}=-\frac{{\boldsymbol\nabla}p^{'}}{{\overline\rho}}+\frac{\rho^{'}}{{\overline\rho}}{\boldsymbol g},
\end{equation}
and continuity equation
\begin{equation}
{\rm D}_{t}\rho^{'}+{\boldsymbol\nabla}\cdot\left({\overline\rho}{\boldsymbol u}\right)=0,
\end{equation}
where $D_{t}=\partial_{t}+{\overline\Omega}\partial_{\varphi}$, $\rho$ and $p$ are the density and the pressure, respectively, and $\vec g$ is the gravity. Each thermodynamic quantity $X$ (density, pressure, entropy, temperature) is expended as $X\left(r,\theta,\varphi,t\right)={\overline X}\left(r\right)+X^{'}\left(r,\theta,\varphi,t\right)$, where ${\overline X}$ is its hydrostatic value and $X^{'}$ its perturbation. For low-frequency IGWs, we can adopt the Cowling approximation \citep{Cowling1941} and neglect the fluctuation of the gravity.

Finally, we introduce the equation for the transport of heat in the adiabatic limit
\begin{equation}
\frac{1}{\Gamma_1}\left[\frac{D_{t} P^{'}}{\overline P}+u_r\frac{{\rm d}\ln{\overline P}}{{\rm d}r}\right]-\left[\frac{D_{t}\rho^{'}}{{\overline\rho}}+u_r\frac{{\rm d}\ln{\overline\rho}}{{\rm d}r}\right]=0,
\end{equation}
where $\Gamma_1=\left(\partial\ln{\overline P}/\partial\ln{\overline\rho}\right)_S$ is the adiabatic exponent at a fixed macroscopic entropy $S$.

Each thermodynamic quantity is expended on spherical harmonics and as Fourier series in time:
\begin{equation}
X'(r,\theta,\varphi,t)=\displaystyle\sum_{l=0}^{\infty}\sum_{m=-l}^{l}\left\{\hat{X}'_{l,m}(r) Y_{l}^{m}(\theta,\varphi)\right\}e^{i\omega_{\rm in}
          t},
\end{equation}
where $\omega_{\rm in}$ is the wave frequency in an inertial reference frame. The wave velocity field is projected on vectorial spherical harmonics as defined in \cite{Rieutord1987}:
\begin{eqnarray}
  \label{eq:5}
  \vec u (r,\theta,\varphi,t)=
 && \sum_{l=0}^{\infty}\sum_{m=-l}^{l}\left\{
  \hat{u}_{r;l,m} (r)\vec R_l^m(\theta,\varphi)+\right.\\ \nonumber
&&{\left. \hat{u}_{H;l,m} (r)\vec S_l^m(\theta,\varphi)+
  \hat{u}_{T;l,m} (r)\vec T_l^m(\theta,\varphi)
\right\}}e^{i\omega_{\rm in}t},
\end{eqnarray}\\
where
\begin{equation}
\label{Eqy_lm}
  \left\{
      \begin{array}{l}
       \vec R_l^m\left(\theta,\varphi\right) = Y_l^m \left(\theta,\varphi\right)\vec{\hat e}_r \hbox{,}\\
         \vec{S}_l^m\left(\theta,\varphi\right) = \vec\nabla_\bot Y_l^m  = \partial_\theta Y_l^m \vec{\hat e}_\theta
+ \displaystyle\frac{1}{\sin\theta}\partial_\varphi Y_l^m \vec{\hat e}_\varphi \hbox{,}\\
  \vec{T}_l^m\left(\theta,\varphi\right) =  \vec\nabla_\bot \times \vec R_l^m =
  \displaystyle\frac{1}{\sin\theta}\partial_\varphi Y_l^m \vec{\hat e}_\theta - \partial_\theta
  Y_l^m \vec{\hat e}_\varphi
      \end{array}
    \right.
\end{equation}
with the orthogonality relationships: ${\displaystyle\int_{\,{\widetilde\Omega}=4\pi}}\vec R_{l_1}^{m_1}\left(\theta,\varphi\right)\cdot\left(\vec R_{l_2}^{m_2}\left(\theta,\varphi\right)\right)^{*}{\rm }d{\widetilde\Omega}=\delta_{l_1,l_2}\delta_{m_1,m_2}$ and ${\displaystyle\int_{\,{\widetilde\Omega}=4\pi}}\vec S_{l_1}^{m_1}\left(\theta,\varphi\right)\cdot\left(\vec S_{l_2}^{m_2}\left(\theta,\varphi\right)\right)^{*}{\rm }d{\widetilde\Omega}={\displaystyle\int_{\,{\widetilde\Omega}=4\pi}}\vec T_{l_1}^{m_1}\left(\theta,\varphi\right)\cdot\left(\vec T_{l_2}^{m_2}\left(\theta,\varphi\right)\right)^{*}{\rm }d{\widetilde\Omega}=l_1\left(l_1+1\right)\delta_{l_1,l_2}\delta_{m_1,m_2}$, where ${\rm d}{\widetilde\Omega}=\sin\theta{\rm d}\theta{\rm d}\varphi$, $\widetilde\Omega$ being the solid angle, and $*$ designates the complex conjugate. We have introduced the usual Kronecker symbol $\delta_{i,j}=1$ if $i=j$ and that vanishes otherwise. Within this framework, prograde waves are such that $m<0$ while retrograde waves are such that $m>0$.\\ 

We obtain the system of equations for the radial functions derived from the projection of the momentum equation on the vectorial spherical harmonics:
\begin{equation}
\left\{
\begin{array}{l}
i\,\bar\rho\,{\widehat\omega}\,{\hat u}_{r;l,m}=-\displaystyle\frac{\mathrm d
  \hat{p}'_{l,m}}{dr}-\hat\rho'_{l,m}\bar g \hbox{,}\\
i\,\bar\rho\,{\widehat\omega}\,{\hat u}_{H;l,m}=-\displaystyle\frac{\hat{p}'_{l,m}}{r} \hbox{,}\\
i\,\bar\rho\,{\widehat\omega}\,{\hat u}_{T;l,m}=0 \hbox{,}
\end{array}
\right.
\label{eq:momentum}
\end{equation}
and from the projection of the equation of mass conservation on spherical harmonics:
\begin{equation}
i\,{\widehat\omega}\,\hat\rho'_{l,m}+\frac{1}{r^2} \frac{\partial}{\partial r}(r^2 \bar\rho {\hat u}_{r;l,m})
-\displaystyle\frac{l(l+1)}{r}\bar\rho{\hat u}_{H,l,m} = 0
 \hbox{,}
 \label{eq:continuity}
\end{equation}
and of the equation for the transport of energy in the adiabatic limit:
\begin{equation}
i\,{\widehat\omega}\,\frac{\hat\rho'_{l,m}}{\bar\rho}=i\,{\widehat\omega}\,\frac{1}{\Gamma_1}\frac{\hat{p}'_{l,m}}{\bar{p}}+\frac{N^2}{\bar g}{\hat u}_{r;l,m}
\hbox{,}
\label{eq:heat}
\end{equation}
where ${\widehat\omega}=\omega_{\rm in}+m{\overline\Omega}$ is the Doppler-schifted frequency. We have introduced the total squared Brunt-Va\"is\"al\"a frequency $N^2$, which is related to the combination of the entropy stratification and of the chemical stratification \citep[][]{Zahnetal1997}. It can be decomposed as:
\begin{equation}
N^2=N_T^2+N_{\mu}^2,
\end{equation}
where
\begin{equation}
N_T^2=\frac{{\overline g}\delta}{H_P}\left(\nabla_{\rm ad}-\nabla\right)
\end{equation}
and
\begin{equation}
N_{\mu}^2=\frac{{\overline g}\phi}{H_P}\nabla_{\mu}.
\end{equation}
We have introduced the pressure height scale $H_P=\vert{\rm d}r/{\rm d}\ln{\overline P}\vert$, the gradients $\nabla_{\rm ad}=\left(\partial\ln{\overline T}/\partial\ln{\overline P}\right)_{{\overline S}={\rm cst}}$, $\nabla=\left(\partial\ln{\overline T}/\partial\ln{\overline P}\right)$, $\nabla_{\mu}=\left(\partial\ln{\overline \mu}/\partial\ln{\overline P}\right)$ with $S$ the macroscopic entropy, $T$ the temperature, $\mu$ the mean molecular weight, and the thermodynamics coefficients $\delta=-\left(\partial\ln{\overline\rho}/\partial\ln{\overline T}\right)_{{\overline P},{\overline \mu}}$ and $\phi=\left(\partial\ln{\overline\rho}/\partial\ln{\overline \mu}\right)_{{\overline P},{\overline T}}$ \citep{MaederZahn1998}.

From now on, we focus on low-frequency IGWs. This means that we consider progressive IGWs and not standing gravity modes. Progressive IGWs are damped because of their thermal damping (or breaking) before being able to reach the second turning point after their excitation (i.e. the turning point close to the centre of the star for a late-type star with an outer convective envelope and the one close to the surface for an early-type star with a convective core). On the one hand, they are thus not able to form standing gravity modes. On the other hand, they deposit angular momentum at the place where they are damped or where they break \citep[e.g.][]{Zahnetal1997}. The analytical expression for the critical frequency ($\omega_c$), below which IGWs are progressive, has been derived in \cite{Alvanetal2015} and is given by:
\begin{equation}
{\omega}_{c}=\left[l\left(l+1\right)\right]^{3/8}\left(\int_{r_{t;1}}^{r_{t;2}}K\frac{N\,N_{T}^{2}}{r^3}{\rm d}r\right)^{1/4},
\end{equation}
where $r_{t;1}$ and $r_{t;2}$ are the two turning points in the considered radiation zone at which $\omega=N$ and $K$ is the heat diffusivity. We thus consider frequencies such that $\omega<\omega_c$. In addition, we know that the radiative damping of progressive IGWs is scaling as $N N_T^2/{\omega}^4$ \citep[e.g.][and Eq. \ref{eq:raddamping}]{Zahnetal1997}. As a consequence, the larger the ratio $N/\omega$, the stronger the damping and the transport of angular momentum triggered by IGWs. In addition, when $N/\omega\!>\!\!>\!1$, the JWKB asymptotic method \citep{FromanFroman2005} can be used to solve the radial ODE (Eq. \ref{eq:TGSspherical}) describing the radial propagation of a progressive IGW of given latitudinal and azimuthal orders $l$ and $m$. As a consequence, by low-frequency waves we mean that we consider progressive IGWs such that $\omega<\omega_c$ and $\omega\!<\!\!<\!N$. In the case of main-sequence stars, these frequency hierarchies also allow us to filter out acoustic waves, which are at higher frequencies, within the framework of the anelastic approximation.

Then, the system given by Eqs. (\ref{eq:momentum}-\ref{eq:continuity}-\ref{eq:heat}) reduces to the Schr\"odinger-like propagation equation \citep[e.g.][]{Press1981,Zahnetal1997}:  
\begin{equation}
\label{eq:TGSspherical}
\frac{\mathrm d^2 \Psi_{l,m}(r)}{\mathrm dr^2} + k_r^2(r) \Psi_{l,m}(r)=0 \hbox{,}
\end{equation}
where
\begin{equation}
  \label{eq:9}
  \Psi_{l,m}=\bar\rho^{\frac{1}{2}}r^2{\hat u}_{r;l,m}\hbox{,}
\end{equation}
and
\begin{equation}
  \label{eq:10}
  k_r(r) = \frac{N}{{\widehat\omega}} k_h\left(r\right)\quad\hbox{with}\quad k_h=\frac{\sqrt{l(l+1)}}{r}.
\end{equation}

When ${\widehat\omega}\!<\!\!<\!N$, on can provide an asymptotic solution to Eq. (\ref{eq:TGSspherical}) using the JWKB solution \citep{FromanFroman2005}:
\begin{equation}
\Psi=\frac{A_{\Psi}}{\sqrt{k_r}}\exp\left[i\,{\varepsilon}\int_{r_0}^{r} k_r {\rm d}r'\right],
\end{equation}
where $A_{\Psi}$ is an amplitude coefficient. The parameter $\varepsilon\equiv\pm 1$ determines if the phase of the wave is propagating outward ($\varepsilon=-1$; the energy then propagates inward; this corresponds to the excitation of IGWs by an external convective envelope in solar-type stars) or inward ($\varepsilon=1$; the energy then propagates outward; this corresponds to the excitation of IGWs by a convective core). Using Eq. (\ref{eq:9}), we obtain the JWKB expression of the radial component of the velocity:
\begin{eqnarray}
\lefteqn{{\widehat u}_{r;l,m}\left(r\right)={\widehat u}_{r;l,m}\left(r_0\right)\left(\frac{{\overline\rho}\left(r_0\right)}{{\overline\rho}}\right)^{1/2}\left(\frac{r_0}{r}\right)^{3/2}\left(\frac{N\left(r_0\right)}{N}\right)^{1/2}}\nonumber\\
&&\times\exp\left[i\,{\varepsilon}\int_{r_0}^{r} k_r {\rm d}r'\right],
\label{eq:urJWKB}
\end{eqnarray}
where $r_0$ is the excitation radius of the considered IGW.
\subsection{Quasi-adiabatic propagation}
As recalled in the introduction, the current way to model the transport of angular momentum and the mixing of chemicals by IGWs in SSE is to take into account their radiative damping. Their viscous damping is generally neglected since the Prandtl number comparing the viscosity ($\nu$) to heat diffusion ($K$) is small in stellar radiation zones \citep[e.g. of the order of $10^{-6}$ in the Solar radiative core;][]{BrunZahn2006}. In this framework, it is assumed that the heat diffusion acts as a first-order perturbation of the adiabatic propagation of IGWs \citep[e.g.][]{Press1981,Schatzman1993,Zahnetal1997}. In this framework, the vertical wave number used in the JWKB regime has a small imaginary part $k_{r}^{\left(1\right)}\equiv \tau$, and the quasi-adiabatic JWKB solution for each thermodynamic quantity or component of the velocity (noted as ${\widehat X}_{\rm QA}$) is expressed as:
\begin{equation}
{\widehat X}_{{\rm QA};l,m}\left(r\right)={\widehat X}_{{\rm Ad};l,m}\left(r\right)\exp\left[\displaystyle{-\frac{\tau_{l,m}\left(r\right)}{2}}\right],
\end{equation}
where ${\widehat X}_{{\rm Ad};l,m}$ is the corresponding adiabatic JWKB value and
\begin{equation}
\tau_{l,m}\left(r\right)=\left[l\left(l+1\right)\right]^{3/2}\int_{r_{0}}^{r} K\frac{N\,N_{T}^2}{{\widehat\omega}^4}\frac{{\rm d}r'}{r'^3}
\label{eq:raddamping}
\end{equation}
is the damping rate.

\section{Nonlinear wave breaking}

As in planetary atmospheres, IGWs can break nonlinearly in
stellar interiors. The two main sources of nonlinear breaking are
IGW-driven shear and convective instabilities \citep[see][respectively]{Press1981,Lindzen1981,Sutherland2001}. 

\subsection{Convective wave breaking}

The first missing piece in stellar modelling is the breaking induced
by the local convective overturning of the stable stratification
(Barker \& Ogilvie 2011). This phenomenon has never been modelled analytically in stellar physics.

To improve the situation, the best strategy is then to turn to
the literature in atmospheric physics. Indeed, this community has
to model small-scale transport and mixing processes such as turbulence
and IGWs in global circulation models (hereafter GCM), which are used
to predict the long-term evolution of the climate. Because of the broad range of space and time-scales that have to be considered, the equations for the small-scale processes cannot be directly solved in the GCM and they have to be parametrised \citep{Ribsteinetal2022}. This
issue is the same we wish to adress in stellar physics where we
want to evaluate the impact of the transport of momentum and
chemicals in SSE codes, which allow us to
compute the long-term structural, chemical and rotational evolution
of stars. In this framework, relevant theoretical formalism
has been developed by \cite{Lottetal2012} and \cite{LottGuez2013} to evaluate the
modification of atmospheric zonal mean flows by convectively excited breaking IGWs. In their model, IGWs excited by convective turbulence, for instance at the top of clouds, non-linearly break that triggers a convective instability, which overturns the stable stratification. After a transient, the instability saturates and \cite{Lottetal2012} and \cite{LottGuez2013} propose a scheme to compute the transported momentum. The great interest of the parametrisation they obtained for the wave-triggered Reynolds stresses is that its prediction has been successfully compared to those observed in-situ in the Earth atmosphere by stratospheric balloons \citep{Lottetal2023}.

This methodology can be applied to the case of stellar interiors. However, it has to be adapted to the spherical geometry of the problem because the \cite{Lottetal2012} and \cite{LottGuez2013} formalism was developed to be used in local Cartesian shallow-water models relevant for the thin atmosphere of the Earth.

\subsubsection{Convective breaking and saturation scheme}

\begin{figure}[h]
    \centering
    \includegraphics[width=\linewidth]{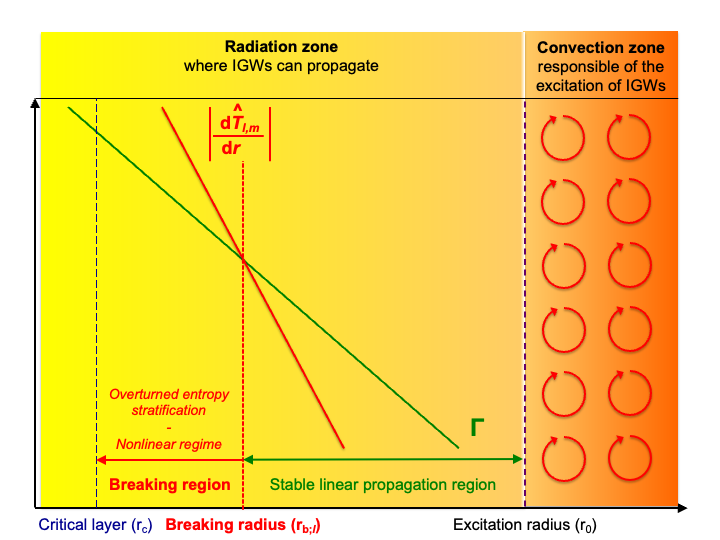}
    \caption{Principle of convective wave breaking: an IGW is excited at $r=r_0$. When propagating towards a critical layer located at $r=r_{\rm c}$, the radial gradient of its temperature fluctuation grows with $k_r$ until it overturns the stable equilibrium entropy stratification at the breaking radius $r=r_b$. This triggers a convective instability.}
    \label{fig:BreakingPrinciple}
\end{figure}

We introduce the macroscopic entropy $S$ definition
\begin{equation}
{\rm d}S=C_{P}\left(\frac{{\rm d}T}{T}-\nabla_{\rm ad}\frac{{\rm d}P}{P}\right),
\end{equation}
where $T$ is the temperature, $C_P$ is the heat capacity at constant pressure and we recall that $\nabla_{\rm ad}=\left(\partial\ln T/\partial\ln P\right)_{S={\rm cst}}$. For low-frequency IGWs, we can assume the anelastic approximation where acoustic waves are filtered out. In this framework, the linearised equation for the transport of heat
\begin{equation}
\rho T {\rm D}_{t}S=\rho T C_{P}\left(\frac{1}{T}{\rm D}_{t}T-\frac{\nabla_{\rm ad}}{P}{\rm D}_{t}P\right),
\end{equation}
reduces to
\begin{equation}
D_{t}T^{'}+\Gamma\,u_{r}=0,\quad\hbox{where}\quad\Gamma=\frac{{\overline T}N_{T}^2}{{\overline g}\delta}.
\label{eq:TempEq}
\end{equation}
When an IGW propagate in the direction of a critical layer (at which ${\widehat\omega}\rightarrow 0$ and $k_{r}\rightarrow\infty$), the gradient of the temperature fluctuation increases (since it scales as $k_r$ in the JWKB regime). At the so-called braking radius $r_{\rm b}$ (derived in the next section), its amplitude reaches the value of the gradient of the background temperature (as represented in Fig. \ref{fig:BreakingPrinciple}):
\begin{equation}
\vert \partial_{r}T^{'}\left(r=r_{\rm b}\right)\vert\approx k_{r}\left(r=r_{\rm b}\right)\vert T^{'}\left(r=r_{\rm b}\right)\vert=\Gamma\left(r=r_{\rm b}\right).
\end{equation}
The breaking IGW thus overturns the stable stratification leading to a convective instability \citep{Lindzen1981,Sutherland2001}. 

At this point, the radial velocity of the IGW reaches its maximum value that we call the saturated value following \cite{Lottetal2012} and \cite{LottGuez2013}. To derive its amplitude, we use the heat transport equation (Eq. \ref{eq:TempEq}) that leads to  
\begin{equation}
\vert u_r\vert=\frac{1}{\Gamma}\vert D_{t}T^{'}\vert.
\end{equation}
Using $\vert D_{t} T'\vert\approx{\widehat\omega}\,\vert T'\vert$ and $\vert\partial_r T'\vert\approx k_r \vert T'\vert$, we obtain:
\begin{equation}
\vert {\widehat u}_{r;l,m} \vert_{\rm sat}=\frac{{\widehat\omega}}{k_r}=\frac{{\widehat\omega}^2}{N k_h}=\frac{1}{\left[l\left(l+1\right)\right]^{1/2}}\left(\frac{{\widehat\omega}}{N}\right)\left(r{\widehat\omega}\right),
\label{eq:satradconv}
\end{equation}
where we have used the IGW dispersion relation $k_r=\left(N/{\widehat\omega}\right)k_h$ with $k_h=\sqrt{l\left(l+1\right)}/r$. We thus recover one of the usual criteria for defining the breaking of waves that states it occurs when the wave velocity reaches the wave phase velocity \citep[e.g.][]{BannerPeregrine1993,SM2022}. Using the asymptotic relation $u_h=\left(k_r/k_h\right)u_r$, that leads to
\begin{equation}
\vert {\widehat u}_{h;l,m} \vert_{\rm sat}=\frac{{\widehat\omega}}{k_h}=\frac{1}{\left[l\left(l+1\right)\right]^{1/2}}\left(r{\widehat\omega}\right).
\end{equation}
for the horizontal component. 

It is interesting to note that these predictions are also in agreement with the criteria for the threshold for IGW nonlinear breaking introduced by \cite{Press1981} who stated that $\left(k_{i}/{\widehat\omega}\right)\,{\widehat u}_{i;l,m}>1$ with $i\equiv\left\{r,h\right\}$.\\

The condition for wave breaking is that $\vert {\widehat u}_{r;l,m} \vert>\vert {\widehat u}_{r;l,m} \vert_{\rm sat}$. Using Eq. (\ref{eq:urJWKB}), this means that the wave have been excited with an initial amplitude such that:
\begin{equation}
\left(\frac{{\overline\rho}\left(r_b\right)}{{\overline\rho}\left(r_0\right)}\right)^{1/2}\left(\frac{r_b}{r_0}\right)^{3/2}\left(\frac{N\left(r_b\right)}{N\left(r_0\right)}\right)^{1/2}\vert {\widehat u}_{r;l,m} \vert_{\rm sat}(r_b)<{\widehat u}_{r;l,m}\left(r_0\right).
\end{equation}
Therefore, for a given value of ${\widehat u}_{r;l,m}\left(r_0\right)$, this equation implicitly determines the value of $r_b$.   

\subsubsection{Saturated vertical flux of energy and angular momentum}
Using Eqs. (\ref{eq:momentum}) and (\ref{eq:heat}) assuming that ${\widehat\omega}\!<\!\!<\!N$ and the anelastic approximation, we express the pressure
fluctuation as a function of the vertical velocity:
\begin{equation}
P^{'}={\varepsilon}\,{\overline\rho}\frac{N^2}{k_r{\widehat\omega}}u_r.
\label{eq:polarisation}
\end{equation}
This allows us to compute the saturated transported monochromatic
flux of energy for convective wave breaking (hereafter CWB):
\begin{eqnarray}
\lefteqn{F_{{\rm E};l,m}^{\rm CWB}\left(r\right)=\frac{1}{4\pi}\int_{{\widetilde\Omega}=4\pi}{\rm Re}\left[P^{'}_{l,m}\left(r,\theta,\varphi,t\right)\right]{\rm Re}\left[u^{*}_{r;l,m}\left(r,\theta,\varphi,t\right)\right]{\rm d}{\widetilde{\Omega}}}\nonumber\\
&=&\!\frac{1}{4\pi}\,\frac{1}{2}\,{\varepsilon}\,{\overline\rho}\,\frac{N^2}{k_r{\widehat\omega}}\left(\frac{{\widehat\omega}^2}{N k_h}\right)^2\!=\!\frac{1}{4\pi}\frac{1}{2}\frac{\varepsilon\,\overline\rho}{\left[l\left(l+1\right)\right]^{3/2}}\left(\frac{\widehat\omega}{N}\right)\left(r{\widetilde\omega}\right)^3,
\end{eqnarray}
where ${\rm d}{\widetilde{\Omega}}=\sin\theta{\rm d}\theta{\rm d}\varphi$ is the elementary solid angle, ${\rm Re}$ is the real part, and $^*$ is the complex conjugate, while $u_{r;l,m}={\widehat u}_{r;l,m}\left(r\right)Y_{l}^{m}\left(\theta,\varphi\right)e^{i\omega_{\rm in}t}$, $P^{'}_{l,m}={\widehat P}_{l,m}\left(r\right)Y_{l}^{m}\left(\theta,\varphi\right)e^{i\omega_{\rm in}t}$ with ${\widehat P}_{l,m}=\varepsilon\,{\overline \rho}\,N^2\,{\widehat u}_{r;l,m}/\left(k_r{\widehat\omega}\right)$ where $\vert {\widehat u}_{r;l,m}\vert=\vert {\widehat u}_{r;l,m} \vert_{\rm sat}$ (Eqs. \ref{eq:polarisation} and \ref{eq:satradconv}, respectively). In the case of a solar-type star with an external convective envelope, we recover that the energy is injected inward with $\varepsilon=-1$. In the case of an early-type star with a convective core, it is injected outward with $\varepsilon=1$. This allows us to derive the triggered flux of angular momentum:
\begin{eqnarray}
\lefteqn{F_{{\rm J};l,m}^{\rm CWB}\left(r\right)=-\frac{m}{\widehat\omega}F_{{\rm E};l,m}^{\rm CWB}\left(r\right)}\nonumber\\
&=&-\frac{1}{4\pi}\frac{1}{2}\frac{m\,{\varepsilon}\,{\overline\rho}}{\left[l\left(l+1\right)\right]^{3/2}}\left(\frac{\widehat\omega}{N}\right)\frac{\left(r{\widehat\omega}\right)^3}{\widehat\omega}.
\end{eqnarray}
When taking this expression and applying it in a local Cartesian
plane, we recover the expression derived by \cite{Lottetal2012} and \cite{LottGuez2013}. We recall the interest of their parametrisation, which has been validated by in-situ measurements of wave-induced Reynolds stresses in the Earth atmosphere by stratospheric ballons \citep{Lottetal2023}. In the case of a solar-type star for which $\varepsilon=-1$, we recover that a retrograde wave ($m>0$) is extracting angular momentum from the radiative core while a prograde one ($m<0$) is depositing angular momentum there. The situation is the opposite in the case of an early-type star for which $\varepsilon=1$.

It is important to examine the physical behaviour of the transported flux of angular momentum. First, as expected, it vanishes when the IGW propagate very close to a critical layer where ${\widehat\omega}\rightarrow 0$. This means that the angular momentum is deposited very efficiently close to this layer. Next, we see that the higher the latitudinal degree $l$, the stronger the deposit. Such a behaviour is similar to the case of the deposit through the radiative damping, which is also more efficient for high-degree IGWs. 

The transported flux diminishes with the strength of the stratification quantified by $N$ that means a more efficient deposit. This can be understood in the following way: at a fixed $\widehat\omega$, the strongest $N$, the strongest the vertical wave number $k_r$ and as a consequence the vertical gradient of the wave triggered temperature fluctuation that overcomes the stable equilibrium entropy gradient. In the specific case where $N\rightarrow 0$, as in the center of solar-type stars, the asymptotic relation $k_r\approx(N/{\widehat\omega})k_h$ cannot be used anymore since we loose the frequency hierarchy $1<\!<\left(N/{\widehat\omega}\right)$ and specific prescriptions must be derived. In this case, the radial buoyancy force vanishes and any motion potentially leads to a convective instability \citep[e.g.][]{BarkerOgilvie2010}.

Finally, a vanishing density ($\overline\rho$) also leads to a flux that vanishes and an efficient deposit of angular momentum. This corresponds for instance to the case of layers close to the surface of early-type stars where such an efficient momentum deposit could drive matter ejection as in active Be stars \citep{Rogersetal2013,Rogers2015,Neineretal2020}.

\subsection{Breaking due to IGWs shear instabilities}
The great interest of the method proposed by \cite{Lottetal2012} and \cite{LottGuez2013} is that it allows us to derive the transported fluxes of energy and angular momentum by IGWs convective breaking by deriving the saturated value of their velocity from the convective instability criteria.

Since IGWs can also break because of the vertical shear instability of their own velocity \citep[][]{GLS1991}, we can apply the \cite{Lottetal2012} and \cite{Lottetal2023} methodology to this instability. 

We first consider the simplest case where the action of the heat diffusion, which weakens the action of the restoring buoyancy force, is ignored. Following \cite{Alvanetal2013}, the Richardson criteria for the vertical shear instability of an IGW is given by:  
\begin{equation}
\frac{N^2}{\left|\displaystyle{\frac{{\rm d}{\sqrt{<{\boldsymbol u}_h\cdot{\boldsymbol u}_h^{*}>_{\theta,\varphi}}}}{{\rm d}r}}\right|^{2}}\le R_{\rm i;c}\,\beta_{l,m},\,\hbox{where}\quad \beta_{l,m}=\frac{m^2}{l\left(l+1\right)}
\label{VSI}
\end{equation}
and $R_{\rm i;c}$ is the critical Richardson number. The geometrical factor $\beta_{l,m}$ comes from the derivation of the Taylor Goldstein Synge equation for IGWs propagating in vertically sheared shellular rotation projected on spherical harmonics in spherical geometry \citep[][]{Alvanetal2013}. Since we consider low-frequency IGWs for which asymptotic JWKB solutions apply, we have $\vert\vert{\boldsymbol u}_r\vert\vert\!<\!<\!\vert\vert{\boldsymbol u}_h\vert\vert$, where ${\boldsymbol u}_h\equiv {\widehat u}_{h;l,m}\left(r\right){\boldsymbol S}_{l}^{m}\left(\theta,\varphi\right)$ is theirs horizontal component; we have introduced the horizontal average over the sphere $<X>_{\theta,\varphi}=\int_{\,{\widetilde \Omega}=4\pi}X\left(\theta,\varphi\right){\rm d}{\widetilde \Omega}$, where we recall that ${\rm d}{\widetilde \Omega}=\sin\theta{\rm d}\theta{\rm d}\varphi$ is the elementary solid angle element. The introduction of this horizontal average is motivated by our goal to evaluate horizontal averages of the transported fluxes of energy and angular momentum. It is thus more appropriate than a purely local criteria that may be only applied at a given co-latitude. Using the JWKB identity ${\rm d}/{{\rm d}r}\propto i\,\varepsilon\,k_r$, Eq. (\ref{VSI}) can be written as:
\begin{equation}
\frac{1}{R_{\rm i;c}\,\beta_{l,m}}\left(\frac{\widehat\omega}{k_h}\right)^2\le\left(\vert {\widehat u}_{h;l,m}\vert_{\rm sat}\right)^2.
\label{VSIuh}
\end{equation}
Using the JWKB polarisation relation $u_h\approx \left(k_r/k_h\right)\,u_r$ with $k_h=\left[l\left(l+1\right)\right]^{1/2}/r$, this leads to the corresponding criterion for the vertical velocity:
\begin{equation}
\frac{1}{R_{\rm i;c}\,\beta_{l,m}}\left(\frac{\widehat\omega}{k_r}\right)^2\le\left(\vert {\widehat u}_{r;l,m}\vert_{\rm sat}\right)^2.
\label{VSIur}
\end{equation}
As in the case of the convective breaking, we recover a criteria involving that the component of the velocity field have to overcome the corresponding component of the wave phase velocity.

These equations allow us to derive the explicit expression for the saturated value for the squared amplitude of the vertical component of IGWs:  
\begin{equation}
\left(\vert {\widehat u}_{r;l,m}\vert_{\rm sat}\right)^2=\frac{1}{R_{\rm i;c}\,m^2}\left(\frac{{\widehat\omega}}{N}\right)^2\left(r{\widehat\omega}\right)^2.
\end{equation}
This allows us to obtain the saturated amplitudes of the flux of energy
\begin{equation}
F_{{\rm E};l,m}^{\rm SWB}=\frac{1}{4\pi}\frac{1}{2}\frac{{\varepsilon}\,\overline\rho}{R_{\rm i;c}\,m^2\,\left[l\left(l+1\right)\right]^{1/2}}\left(\frac{\widehat\omega}{N}\right)\left(r{\widehat\omega}\right)^3
\end{equation}
and angular momentum
\begin{equation}
F_{{\rm J};l,m}^{\rm SWB}=-\frac{1}{4\pi}\frac{1}{2}\frac{m\,{\varepsilon}\,{\overline\rho}}{R_{\rm i;c}\,m^2\,\left[l\left(l+1\right)\right]^{1/2}}\left(\frac{\widehat\omega}{N}\right)\frac{\left(r{\widehat\omega}\right)^3}{\widehat\omega},
\end{equation}
along the radial direction, respectively, in the case of shear-driven wave breaking (hereafter SWB).

Given the similar form of the saturated velocity when compared to the convective breaking, we get the same dependences of the flux of angular momentum as a function of ${\widehat\omega}$, $l$, $N$, and ${\overline\rho}$. We note a supplementary dependence on $m^2$ due to the geometrical factor $\beta_{l,m}$. The largest $m$, the strongest the deposit of angular momentum. Note however, that this dependence on $m^2$ does not introduce a supplementary bias between pro- and retrograde waves. Finally, the deposit of angular momentum is more efficient for high values of the critical Richardson number ($R_{\rm i;c}$). This can be understood by looking at Eq. (\ref{VSI}): an higher $R_{\rm i;c}$ diminishes the instability threshold $N^2/\left(R_{\rm i;c}\beta_{l,m}\right)$.\\ 

However, the action of heat diffusion is of great importance in stellar interiors, because it favours the development of vertical shear instabilities \citep{Townsend1958,Zahn1992,L1999,PratL2013,Pratetal2014}. The vertical shear instability criterion becomes
\begin{equation}
\frac{N^2}{\left|\displaystyle{\frac{{\rm d}{\sqrt{<{\boldsymbol u}_h\cdot{\boldsymbol u}_h^{*}>_{\theta,\varphi}}}}{{\rm d}r}}\right|^{2}}\frac{\sqrt{<{\boldsymbol u}_r\cdot{\boldsymbol u}_r^{*}>_{\theta,\varphi}}\,l_r}{K}\le R_{\rm i;c}\,\beta_{l,m},
\end{equation}
where we identify the Peclet number $P_{\rm e}\equiv\left(\sqrt{<{\boldsymbol u}_r\cdot{\boldsymbol u}_r^{*}>_{\theta,\varphi}}\,l_r\right)/K$, which evaluates the ratio between heat vertical advection and diffusion. At the saturation level for IGWs for which we reach the instability threshold, we can identify the radial characteristic length scale $l_r$ to $1/k_r$. This allows us to derive the scalings for the saturated values of the vertical components of IGWs velocity:
\begin{equation}
\vert {\widehat u}_{r;l,m}\vert_{\rm sat}=\frac{1}{R_{\rm i;c}\,\beta_{l,m}\,\left[l\left(l+1\right)\right]^{3/2}}\left(\frac{r^2\,N}{K}\right)\left(\frac{{\widehat\omega}}{N}\right)^4\left(r{\widehat\omega}\right),
\end{equation}
energy flux
\begin{equation}
F_{{\rm E};l,m}^{\rm SWB}=\frac{1}{4\pi}\frac{1}{2}\frac{\varepsilon\,{\overline\rho}}{R_{\rm i;c}^2\,\beta_{l,m}^{2}\,\left[l\left(l+1\right)\right]^{7/2}}\left(\frac{r^2\,N}{K}\right)^2\left(\frac{\widehat\omega}{N}\right)^{7}\left(r{\widehat\omega}\right)^3
\end{equation}
and angular momentum
\begin{equation}
F_{{\rm J};l,m}^{\rm SWB}=-\frac{1}{4\pi}\frac{1}{2}\frac{m\,{\varepsilon}\,{\overline\rho}}{R_{\rm i;c}^2\,\beta_{l,m}^{2}\,\left[l\left(l+1\right)\right]^{7/2}}\left(\frac{r^2\,N}{K}\right)^2\left(\frac{\widehat\omega}{N}\right)^{7}\frac{\left(r{\widehat\omega}\right)^3}{\widehat\omega}.
\end{equation}
The key modification in this case is the dependence of the saturated velocity and of the transported fluxes on the heat diffusivity ($K$). The strongest the heat diffusion, the more efficient the deposit of angular momentum. This can be understood since the heat diffusion diminishes the stability of a vertically-sheared flow that favours the IGW shear-instability driven breaking. The modified form of the stability equation is the reason of the modification of the dependence of the flux on ${\widehat\omega}$, $l$, $m$, and $N$. If the involved powers are higher, the trends are however the same that in the case where the action of heat diffusion was neglected.

\subsection{Modelling angular momentum transport}

The computed transported flux of angular momentum allows us to evaluate its evolution \citep[e.g.][]{Zahnetal1997,Mathis2009}:
\begin{eqnarray}
\left[{\overline\rho}\frac{{\rm d}\left(r^2{\overline\Omega}\right)}{{\rm d}t}\right]_{\rm IGWs}=-\frac{3}{2}\frac{1}{r^2}\partial_{r}\left[r^2\,F_{\rm J}\left(r\right)\right].
\end{eqnarray}
From the previous section, we know that the flux cannot exceed its value in the saturated regime when IGWs nonlinear breaking occurs. We can thus model the transported flux of angular momentum as:
\begin{eqnarray}
\lefteqn{F\left(r\right)=}\nonumber\\
& &\sum_{\omega_{\rm in},l,m}{\rm min}\left\{-\frac{m}{\widehat\omega}F_{{\rm E};l,m}\left(r_0\right)\exp\left[-\tau_{l,m}\left(r\right)\right]\right.,\nonumber\\
& & -\frac{1}{4\pi}\frac{1}{2}\frac{m\,{\varepsilon}\,{\overline\rho}}{\left[l\left(l+1\right)\right]^{3/2}}\left(\frac{\widehat\omega}{N}\right)\frac{\left(r{\widehat\omega}\right)^3}{\widehat\omega},\nonumber\\
& & {\left.-\frac{1}{4\pi}\frac{1}{2}\frac{m\,{\varepsilon}\,{\overline\rho}}{R_{\rm i;c}^2\,\beta_{l,m}^{2}\,\left[l\left(l+1\right)\right]^{7/2}}\left(\frac{r^2\,N}{K}\right)^2\left(\frac{\widehat\omega}{N}\right)^{7}\frac{\left(r{\widehat\omega}\right)^3}{\widehat\omega}\right\}},
\end{eqnarray}
where the first flux in the bracket corresponds to its standard quasi-adiabatic modelling \citep[e.g.][]{Schatzman1993,Zahnetal1997}. This expression is the generalisation of the expression derived in \cite{Lottetal2012}, \cite{LottGuez2013} and \cite{Ribsteinetal2022} for the studied case of deep spherical stellar regions with taking into account both the convective and vertical shear driven breakings. Its analytical form makes it ready to be implemented in 1D SSE numerical codes.

\section{Taking into account the Coriolis and Lorentz forces}
Early-type stars and young late type stars are rapid rotators \citep[e.g.][]{GalletBouvier2013,AT2023}. In addition, stellar radiation zones can be the seat of magnetic fields (stable or unstable) inherited from previous convective phases where dynamo action has generated magnetic energy \citep[e.g.][]{BS2004,DM2010,Arlt2014,EVB2017,Bugnetetal2021}. As a consequence, low-frequency IGWs that transport angular momentum can be simultaneously modified by the Coriolis acceleration and the Lorentz force and become magneto-gravito-inertial waves \citep[hereafter MGIWs;][]{KTZ1999,MGR2010,MdB2011,MdB2012,Ruietal2024}.\\

In the case where the Archimedean force is larger than the radial components of the Coriolis acceleration and of the Lorentz force, the dynamics of MGIWs can be modelled semi-analytically using the Traditional Approximation for Rotation and Magnetism \citep[hereafter TARM; e.g.][]{MdB2011,Ruietal2024}. 

Let us consider the most advanced semi-analytical model for the transport of angular momentum by MGIWs in stellar radiation zones developed by \cite{MdB2012}. In this model, we consider, as a first step, a rotation profile
\begin{equation}
\Omega\left(r,\theta\right)=\Omega_s+\delta{\Omega}\left(r\right),
\end{equation}
which is the sum of a global solid rotation $\Omega_s$ and of a weak radial differential rotation $\delta{\Omega}$. Next, we choose a purely axisymmetric azimuthal magnetic field defined as in \cite{Malkus1967}:
\begin{equation}
{\boldsymbol B}\left(r,\theta\right)=\sqrt{\mu{\overline\rho}}r\sin\theta\omega_{\rm A}{\widehat{\boldsymbol e}}_{\varphi}
\end{equation}
corresponding to a constant Alfv\'en frequency $\omega_{\rm A}$. If this magnetic topology is a very specific simplified one, it allows us to explore the dynamics of the waves within the radiation zone as function of the frequency hierarchies between $N$, $\Omega_s$, and $\omega_{\rm A}$ when $\left\{\omega_{\rm A},2\Omega_s\right\}\!<\!\!<\!N$ with taking into account both the Coriolis acceleration and the Lorentz force in a non-perturbative way. The interesting point is that more complex models taking into account more general axisymmetric differential rotation and magnetic field \citep[e.g.][]{Dhouibetal2022,Ruietal2024} show similar behavior that this simplified model, which thus constitutes an interesting first laboratory.

Within this simplified model, \cite{MdB2011,MdB2012} have identified a key modified frequency
\begin{equation}
\omega_{\rm M}^2=\omega_{s}^2-m^2\omega_{\rm A}^2
\end{equation}
and the magnetic spin control parameter 
\begin{equation}
\nu_{\rm M}=\frac{2\Omega_s}{\omega_s}\frac{1-m\displaystyle{\frac{\omega_{\rm A}^2}{\Omega_{s}\omega_{s}}}}{1-\displaystyle{\frac{m^2}{2}\frac{2\Omega_s}{\omega_s}\frac{\omega_{\rm A}^2}{\Omega_{s}\omega_{s}}}},
\end{equation}
which allow us to model the propagation of MGIWs in a theoretical formalism similar to the non-rotating non-magnetic case. The only difference is that the associated Legendre polynomial of the spherical harmonics is replaced by a so-called Hough function $\Theta_{k,m}\left(\cos\theta;\nu_{\rm M}\right)$ with an associated horizontal eigenvalue $\Lambda_{\k,m}\left(\nu_{\rm M}\right)$ that reduces to $l\left(l+1\right)$ when $\Omega_s$ and $\omega_{\rm A}$ vanish. We refer the reader to \cite{Hough1898}, \cite{LH1968}, \cite{LeeSaio1997}, and \cite{Townsend2003,Townsend2020} for their detailed properties. In this framework, the MGIWs JWKB dispersion relation becomes
\begin{equation}
k_r=\left(\frac{N}{\omega_{\rm M}}\right)k_h\quad\hbox{with}\quad k_h=\frac{\sqrt{\Lambda_{k,m}\left(\nu_{\rm M}\right)}}{r}.
\end{equation}
The place where $\omega_{\rm M}$ vanishes corresponds to a magnetic critical layer where MGIWs become trapped by the magnetic field and cannot propagate anymore \citep[e.g.][]{Schatzman1993Mag,MGR2010,MdB2011}. As a consequence the transport of angular momentum by MGIWs vanishes.\\

In the following, we will focus on the impact of the Coriolis and Lorentz forces on the convective breaking of MGIWs. Indeed, to our knowledge, there is no available general criterion for the instability of a radial shear in presence of both rotation and magnetic field \citep[e.g.][]{SalhiCambon2010,Lecoanetetal2010}. Inserting the dispersion relation for MGIWs into the heat transport equation, where we neglect the effects of the Ohmic heating, the amplitude of the saturated velocity becomes:
\begin{equation}
\vert {\widehat u}_{r;k,m} \vert_{\rm sat}=\frac{{\widehat\omega}\,\omega_{\rm M}}{N k_h}\quad\hbox{with}\quad k_h=\frac{\sqrt{\Lambda_{k,m}\left(\nu_{\rm M}\right)}}{r}.
\end{equation}
Using the polarisation relation (Eq. \ref{eq:polarisation}) within the TARM, we derive the monochromatic transported flux of angular momentum:
\begin{equation}
F_{{\rm J};k,m}^{\rm CWB}=-\frac{1}{2}\,\frac{1}{2}\,m\,{\varepsilon}\,{\overline\rho}\,r \frac{\omega_{\rm M}^3\, r^2}{N \Lambda_{k,m}^{3/2}\left(\nu_{\rm M}\right)}\left<\Theta_{k,m}^{2}\left(\cos\theta;\nu_{\rm M}\right)\right>_{\theta},
\end{equation}
where $\left<\cdot\!\cdot\!\cdot\right>_{\theta}=\int_{0}^{\pi}\cdot\!\cdot\!\cdot\sin\theta{\rm d}\theta$. Since the horizontal eigenvalue $\Lambda_{k,m}$ is larger than $l\left(l+1\right)$ while $\left<\Theta_{k,m}^{2}\left(\cos\theta;\nu_{\rm M}\right)\right>_{\theta}$ is smaller than the latitudinal average of the corresponding associated Legendre polynomial when the rotation and the magnetic field vanish, $F_{\rm J;breaking}$ diminishes in presence of rotation and magnetism. This can be understood since we are here considering the convective overturning of the stable stratification while rotation and magnetic field are known to potentially decrease the efficiency of convection for its mode transporting the most heat \citep[e.g.][]{Stevenson1979,AugustsonMathis2019}. When $\omega_{\rm M}$ vanishes, meaning that the IGW meets a critical magnetic layer, we recover that the flux of angular momentum vanishes. 

Following the same methodology as in the non-rotating and non-magnetic case, we finally obtain the parametrisation of the angular momentum tranport:
\begin{eqnarray}
\lefteqn{F\left(r\right)=\sum_{\omega_{\rm in},k,m}{\rm min}\left\{-\frac{m}{\widehat\omega}F_{{\rm E};k,m}\left(r_{0}\right)\exp\left[-\tau_{k,m}\left(r\right)\right]\right.}\nonumber\\
& & {\left. ,-\frac{1}{2}\,\frac{1}{2}\,m\,{\varepsilon}\,{\overline\rho}\,r \frac{\omega_{\rm M}^3\, r^2}{N \Lambda_{k,m}^{3/2}\left(\nu_{\rm M}\right)}\left<\Theta_{k,m}^{2}\left(\cos\theta;\nu_{\rm M}\right)\right>_{\theta}\right\}},
\end{eqnarray}
where we recall the value of the thermal damping in the rotating magnetic case as derived in \cite{MdB2012}:
\begin{equation}
\tau_{k,m}=\Lambda_{k,m}^{3/2}\left(\nu_{{\rm M};m}\right)\int_{r}K\frac{N_T^2\,N}{{\widehat\omega}\,\omega_{\rm M}^3}\frac{d r'}{r'^{3}}.
\end{equation}
\section{Conclusion and perspectives}
In this work, we have studied the nonlinear breaking of IGWs and the transport of angular momentum it triggers. We have taken advantage of theoretical progresses for its modelling in the Earth atmosphere, which have been achieved ten years ago in local Cartesian models \citep{Lottetal2012,LottGuez2013}, implemented in global circulation models \citep[e.g.][]{Ribsteinetal2022} and whose predictions have been successfully compared to in-situ measurements of Reynolds stresses triggered by IGWs by stratospheric balloons \citep{Lottetal2023}. 

First, we have considered the non-rotating and non-magnetic case. We have identified the condition in which the temperature fluctuation of an IGW is able to overturn the surrounding stable stratification or its velocity becomes unstable because of its vertical shear. We derived the IGWs amplitude for which convective or vertical shear instabilities occur and the corresponding breaking radius even if we should remember that once turbulence is developing, such a local criterion is an approximation of the reality. From the instabilities criteria, we derived the values of the saturated wave velocity necessary to trigger them. This allows us to derive the value of the transported flux of energy and angular momentum. We show that the deposit of angular momentum by breaking waves is increasing with their latitudinal degree and the ratio of the Brunt-Va\"is\"al\"a frequency and the wave frequency and when the density decreases or the Doppler-shifted frequency vanishes. 

As in Geophysics, we thus obtain a theoretical semi-analytical prescription for the wave-mean differential rotation interaction, which takes into account simultaneously the smooth thermal damping of IGWs and their nonlinear breaking. In this vision, IGWs break before reaching critical layers and this is the reason why the prescription derived in \cite{Alvanetal2013} is not taken into account. Such a prescription thus allows us to bring the modelling of the transport of angular momentum by waves over long timescales in stellar structure and evolution numerical codes to the same level of realism that in Global Circulation Models for the Earth atmosphere and the prediction of the evolution of the climate.

In a second time, we have considered the case of rotating magnetised stellar radiation zones in which the non-perturbative actions of the Coriolis acceleration and of the Lorentz force on IGWs must be taken into account. Assuming simplifying assumptions of a weak radial differential rotation and an azimuthal magnetic topology that corresponds to a constant Alfv\'en frequency, we generalised the results we have obtained in the case where the modification of IGWs by the Coriolis acceleration and the Lorentz force are ignored. We consider the regime of the Traditional Approximation for Rotation and Magnetism (TARM) where the radial component of the Coriolis acceleration and of the Lorentz force are dominated by the buoyancy Archimedean force. In this case, a semi-analytical approach is still possible and we see that the transported flux of angular momentum by IGWs convective breaking will diminish when compared to the non-rotating and non-magnetic case. This can be understood as a consequence that rotation and magnetism can inhibit the efficiency of convection for its mode transporting the more heat.

The perspectives of this work can be structured along three axis. The first one will be to implement the obtained prescription in stellar structure and evolution numerical codes, where physical prescriptions for the transport of angular momentum are implemented, and to explore different stellar masses and evolutionary stages \citep[e.g.][]{CharbonnelTalon2005,TalonCharbonnel2005}. Next, more general differential rotation \citep{Mathis2009,Dhouibetal2021} or magnetic topologies \citep{Dhouibetal2022,Ruietal2024} can be considered. For instance, asteroseismology is not providing us yet any constrain on the latitudinal variation of differential rotation in stellar radiation zones, while their interfaces with adjacent convective regions (i.e. stellar tachoclines) can be the seat of strong horizontal shears because of the transition between a differential rotation that varies both in radius and latitude in convection zones and a weaker differential rotation in stellar radiative regions \citep[e.g.][]{SpiegelZahn1992,Brunetal2005,Brunetal2011,Brunetal2017,Brunetal2022}. Moreover, the 2D stellar structure and evolution code ESTER predicts a possible 2D angular velocity profile in the external radiative envelope of rapidly-rotating early-type stars \citep[e.g.][]{ELR2013,Rieutordetal2016,MREL2023}. In these configurations within the framework of the TARM, generalised Hough functions describing the latitudinal behaviour of IGWs replace the standard Hough functions but the main trends described here with our simplifying assumptions should be conserved. Finally, because of the rapid rotation of young late type stars and the rapid rotation and weaker stratification in early type stars it will be necessary to go beyound the TARM and to take into account the full Coriolis acceleration and Lorentz force \citep[e.g.][]{DR2000,GerkemaShrira2005,MNT2014,Mirouhetal2016}.

\begin{acknowledgements}
S.M. thanks the referee for her/his careful reading of the manuscript and for her/his constructive and precise comments, which have allowed him to improve the initial version of the article. S.M. acknowledges support from the European Research Council (ERC) under the Horizon Europe programme (Synergy Grant agreement 101071505: 4D-STAR), from the CNES SOHO-GOLF and PLATO grants at CEA-DAp, and from PNPS (CNRS/INSU). While partially funded by the European Union, views and opinions expressed are however those of the author only and do not necessarily reflect those of the European Union or the European Research Council. Neither the European Union nor the granting authority can be held responsible for them.
\end{acknowledgements}

%
%

\bibliographystyle{aa}
\bibliography{Breaking_Mathis2024_Arxiv}

\end{document}